\documentstyle[aps,12pt]{revtex}

\begin{document}
\title{Gottfried and Bjorken integrals \\ and \\ gluon polarization in the 
proton \\ Talk given at "Trends in Collider Spin Physics", \\ 
Trieste, 5-8 December 1995}   
\vskip 0.30in
\author{E. Di Salvo}
\address{Dipartimento di Fisica and I.N.F.N. - Sez. Genova,
Via Dodecaneso,33 - 16146 Genova, Italy}
\date{\today}
\maketitle
\vskip 0.30in
~~~ \ ~~~~~ \ ~~~~~~ \ ~~~~~~~~ \ ~~~~ \ ~~~~~ \ GEF - Th - 2/1996
\vskip 0.30in
\begin{abstract}
Abstract - We show that the Gottfried and the Bjorken integrals have the same
nonperturbative evolution, which is related to the gluon polarization in the
proton.
\end{abstract}
\vskip 0.25in
\baselineskip=1.2\baselineskip
In the present talk we focus our attention on the Gottfried integral 
and on the Bjorken integral, two quantities which are deduced from data of 
deep inelastic scattering of leptons on nucleons and which are of great 
theoretical interest\cite{refG,refB}. Starting from some assumptions on 
nonperturbative processes, we show that at low $Q^2$ these integrals have 
the same QCD evolution.
Furthermore we show that the two integrals are related to the gluon
polarization inside the proton, an intriguing quantity to resolving the 
so-called "spin crisis".

First of all we recall the definitions of the two above mentioned integrals,

\begin{equation}
S_G (Q^2) = \int_0^1 dx [u(x,Q^2) - d(x,Q^2)], \ ~~ \ ~~ \
S_B (Q^2) = \int_0^1 dx [\Delta u(x,Q^2) - \Delta d(x,Q^2)], 
\end{equation}

and of other integrals which, as we shall see, result to be related to them,
that is

\begin{equation}
\Delta G (Q^2) = \int_0^1 dx \Delta g(x,Q^2), \ ~~ \ ~~ \
\Delta \Sigma (Q^2) = \int_0^1 dx \sum^3_{i=1} \Delta q_i (x,Q^2), 
\label{g4}
\end{equation}

where

\begin{equation}
q_i (\Delta q_i) = q^+_i \pm q^-_i + {\overline q^+_i} \pm {\overline q^-_i}  
\ ~~~ \ (q_1 = u, \ ~ \ q_2 = d, \ ~ \ q_3 = s), \ ~~~  \Delta g = g^+ - g^- .
\end{equation}

According to standard notations $q$, $\overline q$ and $g$ are the quark, 
antiquark and gluon densities,
indices $+$ and $-$ referring to parton helicities. Moreover for the quark spin 
content $\Delta \Sigma$ in the proton we have adopted a 
gauge invariant definition\cite{refCH}.

The QCD evolution properties of the Gottfried and of the Bjorken integrals have 
to be examined in the framework of the hadron structure, which appears different
according as we consider masses and magnetic moments, or
quantities inferred from deep inelastic scattering, weak and pionic
decays. As is well-known, hadronic masses and magnetic moments are described 
satisfactorily by a nonrelativistic constituent quark model (NRCQM), so that 
the hadronic states constitute irreducible representations of the $SU(6) 
\otimes O(3)$ group. On the contrary deep inelastic scattering (DIS) and decays
support the current quark picture, with respect to which
hadrons appear as complicated mixings of irreducible representations of that
group. In particular according to the NRCQM the nucleon belongs to the 
irreducible representation $\{56, L=0\}$. Therefore we predict that

\begin{equation}
S_G = S_G^0 = 1; \ ~~~ \ ~~~ \ S_B = S_B^0 = \frac{5}{3}; \ ~~~ \ ~~~ \ 
\ \Delta \Sigma = 1 .
\label{int1}
\end{equation}

The parton model is not expected to agree with the
predictions  (\ref{int1}), owing to the above mentioned mixing. Indeed several 
years ago three sum rules - by Gottfried, Bjorken and Ellis-Jaffe - were 
formulated about the above mentioned integrals. Among these only the 
Gottfried sum rule
\cite{refG}, based on the assumption of isoscalar sea, was in accord with 
(\ref{int1}). The Bjorken sum rule\cite{refB}, founded on the 
parton model and on isospin symmetry, predicted that $S_B \simeq 1.26$; 
moreover the Ellis-Jaffe sum rule\cite{refEJ}, based on the assumptions of
unpolarized  
strange sea ($\Delta s(x) = 0$), and on $SU(3)$ symmetry among 
hyperon decay constants, led to the prediction $\Delta \Sigma \simeq 0.7$ 
Surprisingly enough, experiments of a few years ago  have not
confirmed the Gottfried sum rule and, above all, the Ellis-Jaffe sum rule. The 
EMC\cite{refEMC} experimental data, combined with $SU(3)$ flavour symmetry, led 
to the results $\Delta \Sigma = 0.31 \pm 0.07$, $\Delta S < 0$, 
where $\Delta S$ is the spin content of the strange sea. This unexpected result
is usually known as the "spin crisis". Furthermore NMC\cite{refNMC} experiments 
yielded $S_G = 0.72 \pm 0.048$.
Only the Bjorken sum rule (based on very general assumptions) is confirmed by 
experiments\cite{refEXP}, provided perturbative QCD\cite{refKO} corrections are 
taken into account. In any case we conclude that NRCQM
predictions of the quantities $S_G$, $S_B$ and $\Delta \Sigma$ are always in
disagreement with experimental results based on deep inelastic scattering. We 
interpret these discrepancies as effects of the $Q^2$-evolution.
In particular, as regards the Gottfried and Bjorken integrals, 
from the data and predictions exposed just above it follows

\begin{equation}
\frac{S_G}{S_G^0}  \sim \frac{S_B}{S_B^0}  \sim 0.75,
\label{fraz}
\end{equation}

where $S^0_{G(B)}$ refer to NRCQM predictions, $S_G$ to DIS measurements and
$S_B$ is the r.h.s. of the Bjorken sum rule.

The approximate equality (\ref{fraz}) suggests that, at sufficiently low $Q^2$ 
(less than $\sim 4$ $GeV^2$), $S_G$ and $S_B$ could have the same 
$Q^2$-evolution, 
caused by nonperturbative interactions. In other words we 
assume an evolution equation similar to the Altarelli-Parisi equation, that is,

\begin{equation}
S(Q^2)  = S(Q^2_0) exp[ \int_{t_0}^t  \gamma (t') dt' ] ,  \ ~~ \ ~~ \ 
\ t =  log \frac{Q^2}{\mu^2} ,
\label{evol}
\end{equation}

where $S$ denotes either the Gottfried or the Bjorken integral
and $\gamma$ a function of $t$ which depends on the 
nonperturbative interactions. We identify the cause of this evolution with
Spontaneous Chiral Symmetry Breaking (SCSB). Let us illustrate in detail the
case of the Bjorken integral. To the leading twist this integral is related to 
the isovector axial current:

\begin{equation}
S_B s_{\mu} = <P,s|j^3_{5\mu}|P,s> C(Q^2), \ ~~ \ ~~ \ 
\  s_{\mu} = <P,s|\gamma_5 \gamma_{\mu}|P,s> ,
\label{axi}
\end{equation}

where $s$ is the proton spin and $C$ a real positive function of $Q^2$, which,
according to the Bjorken sum rule, satisfies the condition
$\displaystyle\lim_{Q^2\to\infty}{C(Q^2)}=1$; 
that is, an infinite momentum probe "sees" the current quarks as noninteracting 
and therefore, as illustrated in fig. 1, the axial charge is the same as the 
one measured in beta-decay. At high but finite $Q^2$ the current quarks 
interact perturbatively through gluons and, according to the results by 
Kodaira et al.\cite{refKO}, cause a reduction of $C(Q^2)$, i. e., of the 
effective axial charge "seen" by virtual photons. However terms of order 
$\alpha_s^2$ tend to increase  the effective
charge. When $Q^2$ becomes sufficiently small, nonperturbative processes have 
to be taken into account and increasing effects prevail over reduction ones; 
in particular SCSB converts 
current quarks into constituent quarks, so as to increase  $C(Q^2)$,  
therefore $S_B$ varies from $\sim 1.26$ to $\sim 1.67$, according to the NRCQM.
 
Now we show that according to the chiral model by Ball and Forte\cite{refBF} 
(named BF in the following) the Gottfried and the Bjorken integrals have the 
same evolution for small $Q^2$. The model is based on nonperturbative 
interactions between quarks and pseudoscalar mesons. 
These interactions, pictured in fig. 2, are dominant at sufficiently low $Q^2$ 
and 
cause an evolution of the quark densities inside the proton, similarly to the 
Altarelli-Parisi perturbative splitting functions. In the BF model the function 
$\gamma$, which appears in the first eq. (\ref{evol}), is given by
$\gamma = \frac{d}{dt} (\frac{1}{2}  \sigma_{\gamma^* \pi^0} + 
\ \frac{1}{3}  \sigma_{\gamma^* \eta} + \frac{1}{6}  \sigma_{\gamma^* \eta'} - 
\  \sigma_{\gamma^* \pi^+})$,
where the $\sigma$ are the cross sections for the processes
$\gamma^* q \longrightarrow q' M$, ($M = \pi, \eta, \eta'$)  
and $t$ is the evolution parameter defined by the second eq. (\ref{evol}). 
The axial anomaly produces differences among the meson masses, which in turn
cause the coefficient $\gamma$ not to vanish. Indeed, the largest
contribution to $\gamma$ comes from the graph containing the pion. BF show that 
this
coefficient is particularly important for not too small $Q^2$, while for $Q^2$
comparable with the mass squared of the pion it is negligibly small. Then the 
chiral approximation may be adopted. In 
the rest frame of a pion the quark and the antiquark have equal helicities;
since for a massless quark the helicity is independent of the reference frame,
the helicity of the final quark in the splitting function is equal to that
of the initial quark (see fig. 2). This implies, similarly to the
Altarelli-Parisi splitting functions, that polarized and unpolarized nonsinglet
structure functions have the same evolution. So the approximate equality 
(\ref{fraz}) follows from the BF model.

Now we examine the question from a complementary viewpoint. The components of 
the hadronic tensor may be either symmetric or antisymmetric, according 
as we consider unpolarized deep inelastic scattering cross section or asymmetry 
with polarized target and beam. Furthermore 
we distinguish between the isovector and the isoscalar components of the
hadronic tensor. In the present talk we are interested only in the isovector
components: the (anti-)symmetric component is proportional to $S_G(S_B)$.  
Fig. 1 
represents the antisymmetric isovector component of the hadronic tensor, that 
is, the forward Compton scattering with the exchange of an isovector 
pseudovector object. This
exchange is described by a meson Regge trajectory. Indeed, since the
antisymmetric isovector component of the hadronic tensor is proportional to 
the isovector axial current, the Goldberger-Trieman relation implies the 
dominance of the pion Regge trajectory.
In principle also the symmetric part of the isovector hadronic tensor 
could be dominated by a Regge trajectory with the right 
quantum numbers, i. e. by the exchange of a scalar isovector 
object. However SCSB removes the degeneracy between the pion and the 
$\rho$-meson Regge trajectories, so that no
simple reggeized meson exchange dominates the isovector component of 
unpolarized forward Compton scattering. A more complex object has to be
hypothesized.
In particular, as illustrated in fig. 3, such an object could 
be constituted by the Lorentz invariant product of the isovector axial current 
times the Chern-Simons current, defined as
$k_{\mu} = \frac{\alpha_s}{2\pi} \epsilon_{\mu \nu \lambda \sigma} 
\ Tr[A^{\nu} (G^{\lambda \sigma} - \frac{2}{3} A^{\lambda} A^{\sigma})],$
where $A$ is the gluon field and $G$ the QCD strength tensor field. Therefore 
we set 

\begin{equation}
S_G = F C(Q^2) <P|j^3_{5\mu} k^{\mu}|P>,
\label{ass3}
\end{equation}

where the constant $F$, $Q^2$-independent, has been chosen in such a way that
$\displaystyle\lim_{Q^2\to0}{S_G}=1.$ 
Indeed the operator $j^3_{5\mu} k^{\mu}$ has the right quantum numbers for
being exchanged in the isovector symmetric hadronic tensor, furthermore it does 
dominate in the forward Compton scattering, nor do we know an equally simple
operator which may have a comparable weight in such a process.

In order to draw the consequences of assumption (\ref{ass3}), we recall some
important properties of the Chern-Simons current, which plays an essential role
in two important effects, like
the $\eta'$-particle mass shift with respect to the pseudoscalar meson 
octet \cite{refTH} (fig. 4) and the "spin crisis"\cite{refCH}.

In connection with the latter effect, we precise that the forward SU(3)-singlet
axial form factor $\Delta \Sigma$ consists in two addends, called
respectively connected ($\Delta \Sigma'$) and disconnected ($- \Delta \Gamma$)
insertions\cite{refCH} (fig. 5), i. e.,

\begin{equation}
\Delta \Sigma = \Delta \Sigma' - \Delta \Gamma
\label{spin}
\end{equation}

Furthermore, $\Delta \Sigma$ corresponds to a gauge-invariant, $Q^2$-dependent
definition of the quark spin content inside the proton, that we have adopted
before (see eq. (\ref{g4})). On the contrary $\Delta \Sigma'$ defines the same
quantity in a gauge-variant, $Q^2$-independent way. $\Delta \Gamma$ , which 
represents the spin screening contribution of the sea quarks, caused by
photon-gluon scattering (fig. 6), is related to the Chern-Simons current:

\begin{equation}
<P,s|k_{\mu}|P,s> = c_0 s_{\mu} \Delta \Gamma,
\label{chern}
\end{equation}

where $c_0$ is a $Q^2$-independent positive constant. Taking into account 
equations (\ref{axi}), (\ref{ass3}) and (\ref{chern}) yields 

\begin{equation}
S_G = - c_0 F S_B \Delta \Gamma,
\label{imp}
\end{equation}

which implies that the product $F ~ \Delta \Gamma$ must be negative.
If, as we are going to show, $\Delta \Gamma$ is very slowly $Q^2$-dependent,
eq. (\ref{imp}) implies that $S_G$ and $S_B$ have approximately the same 
$Q^2$-dependence, which proves the approximate equality (\ref{fraz}).

For sufficiently large $Q^2$ $\alpha_s$ is small and $\Delta \Gamma$ evolves
like $\alpha_s^2$. For smaller $Q^2$ (of order 1 $GeV^2$) the perturbative
evolution no longer holds true and only nonperturbative models, like the BF
model, can be invoked. In the BF model the evolutions of the quantities $\Delta 
\Sigma$ and $\Delta G$ are controlled by splitting functions in which only the 
$\eta'$ particle is involved (fig. 7), therefore the QCD evolution of such
quantities is negligible in
comparison with that of $S_B$ (see also ref. \cite{refGE}). Since 
$\Delta \Sigma'$ is
$Q^2$-independent, from eq. (\ref{spin}) it follows that also $\Delta \Gamma$
has a negligible evolution. 

We conclude observing that the quantity $\Delta G$ is crucial in resolving the 
"spin crisis". Presently three possible scenarios may be assumed:

~~i) $\Delta \Sigma' \simeq 0.7$, according to the Ellis-Jaffe sum rule, which
implies $\Delta G \simeq 2$;

~ii) $\Delta \Sigma' \simeq \Delta \Sigma \simeq 0$, according to chiral
models (e. g. \cite{refFR}), which implies a very small value of $\Delta G$ 
($\leq 0.15$);

iii) $\Delta G < 0$, according to considerations about interactions between
constituent quarks and gluons\cite{refRJ}. In this case the constant $F$, 
which appears in eqs. (\ref{ass3}) and (\ref{imp}), is positive, while in cases
i) and ii) it is negative.

Several experiments have been suggested for distiguishing among the above 
mentioned
scenarios\cite{refF2}. Furthermore data form experiment FNAL E581\cite{refFN}, 
although affected by large uncertainties, indicate a small, positive $\Delta G$.

~~~~~~~~~~~~~~~~~~

\hbox{FIGURE CAPTIONS \hfill}
\noindent

[Fig.1] - The antisymmetric isovector component of the hadronic tensor compared
with the amplitude of beta-decay

[Fig.2] - Splitting functions: a) Altarelli-Parisi graphs;
b) quark-meson interactions. Arrows indicate helicities.

[Fig.3] - The symmetric isovector component of the hadronic tensor according to
our assumption.

[Fig.4] - Coupling scheme proposed by 'tHooft \cite{refTH} for explaining the 
mass shift of the $\eta'$ particle.

[Fig.5] - The forward SU(3)-singlet axial form factor: a) connected insertion;
b) disconnected insertion

[Fig.6] - Quark spin screening produced by photon-gluon scattering.

[Fig.7] - Splitting function for evolution of gluon densities and of
SU(3)-singlet combination of quark densities.

\end{document}